\documentstyle[aps,prb,epsfig]{revtex}

\begin{document}

\twocolumn[\hsize\textwidth\columnwidth\hsize\csname@twocolumnfalse\endcsname
\draft

\title{Pairing-fluctuation effect in quasi-two-dimensional superconductivity}

\author{Xin-Zhong Yan$^*$}

\address{Institute of Physics, Chinese Academy of Sciences, P. O. Box 603, Beijing 100080, China} 

\date{\today}

\maketitle

\widetext
\begin{abstract}
We present a self-consistent approach to deal with the pairing-fluctuation effects in quasi-two-dimensional superconducting systems. Besides the Cooper pairs in the Bose-Einstein condensate, there are pairs occupying the excited states, which results in the predominant fluctuations. The low-lying excited states are the collective modes. On the basis of ladder-diagram approximation, we treat the single particles and the pairs on an equal-footing manner. The Green's function of single particles is obtained as an analytic solution to a cubic equation. The bosonic degrees of freedom are relevant to the pseudogap physics in the high-$T_c$ cuprates. The superconducting order parameter and the transition temperature are substantially reduced from the values of the mean-field theory. The calculated phase boundary of superconducting state can reasonably describe the experiment data for cuprates. 
\end{abstract}

\vfill
\narrowtext
\vskip2pc] 

\section {INTRODUCTION}

The high-temperature superconductors are typical quasi-two-dimensional (quasi-2D) electron systems. The motion of the electrons is mainly confined in the Copper-Oxygen planes with weak coupling between them. In the quasi-2D systems, the fluctuation effects may take important role for describing the superconductivity. As stated by the Mermin-Wagner-Hohenberg (MWH) theorem,$^1$ there is no superconductivity at finite temperature in systems of dimensions less than or equal to 2. This is because that pairing fluctuations prohibit the electrons from coherent pairing. It is therefore easy to understand why the fluctuation effects can be significant in quasi-2D systems. 

There are a number of approaches dealing with the fluctuation effects.$^2$ For the under doped high-$T_c$ cuprates (HTC), Emery and Kivelson have argued that the long-range classical phase fluctuation of the order parameter can significantly suppress the transition temperature $T_c$.$^3$ Above $T_c$, pairing becomes locally without long-range phase coherence, especially at the strong-coupling regime. On the other hand, the superconductivity can be viewed as a consequence of Bose-Einstein condensation (BEC) of the preformed pairs.$^{4-6}$ Along with this approach, much effort has been devoted to investigation of the crossover from the weak-coupling BCS superconductivity to the BEC of bound pairs.$^{5-18}$ Most of the works have been performed for the $s$-wave pairing because of its computational simplicity. 

Here, we note that according to the general theory by Goldstone, Salam, and Weinberg,$^{19}$ there exist Goldstone modes in the broken-symmetry state of systems without long-range interactions. Since the Goldstone modes are the lowest excited states for the pairs, the excitations of pairs to these states are the most predominant fluctuations in the system. Before the Goldstone-Salam-Weinberg theory, Anderson had studied the collective modes in the superconducting state with perturbation treatment.$^{20}$ These collective modes are actually the Goldstone modes. The significance of collective modes has been observed in the physical properties of other systems. An example is the spin-wave theory for the Hubbard model at half filling.$^{21-23}$ The collective modes (spin waves) not only correct the ground-state energy of mean-field theory (MFT), but also modify the order parameter significantly. For the one-dimensional Hubbard model, the ground-state energy by the spine-wave theory is very close to the exact result; the typical error is about $1\%$, but the error of the MFT is about $28\%$ at strong-coupling limit.$^{22}$ For the 2D system at the ground state, the magnitude of the order parameter at strong coupling is reduced to about $60\%$ of the MFT value.$^{21,23}$  Along with the perturbation approach, a number of works have investigated the pairing fluctuation effects.$^{9, 24}$ 

However, for systems of dimensions $\leq 2$ at finite temperature, such a perturbation treatment is not valid as it contradicts with the MWH theorem. Even for quasi-2D systems at finite temperature, especially at strong coupling regime, the perturbation treatment is not accurate. The reason is that the fluctuations and the mean-field ordering are equally significant. Because the Goldstone modes coexist with the coherent pairing and may not be regarded as perturbation in quasi-2D systems, an equal-footing treatment of the single particles and the collective modes is therefore desirable.

In this paper, we investigate the superconductivity in the tight-binding model with $d$-wave attraction. The pairing fluctuations above the mean field stem from the excited pairs. According to the many-particle physics, we need to find out the bosonic Green's function for the pairs. The bosonic Green's function is generally coupled with that of the single particles. We present a formalism treating the single particles and the pairs on the equal-footing manner. In the present approximation, the Green's function for the pairs is described by the symmetrical ladder diagrams. All the Green's functions are self-consistently determined by a number of coupled integral equations. In the BEC picture of the superconductivity, the BEC from single pairs begins to occur at $T_c$. Below $T_c$, with the condensation taking place, the single pairs become moving collectively. Even at the ground state, there remains the zero-point motion. These bosonic degrees of freedom are relevant to the pseudogap physics in cuprates.$^{25,26}$ We calculate the phase boundary of superconductivity and compare the result with experiment data of HTC. Some results of this work have been presented in Ref. 27

\section {HAMILTONIAN}

For describing the electrons, we consider the Hamiltonian
$$
H = \sum_{k\alpha}\xi_k c_{k\alpha }^{\dagger }c_{k\alpha }+
{\frac{1}{N}}\sum_{kk'q}v_{kk'}p^{\dagger }(k,q)p(k',q) \eqno(1)
$$
where $c_{k\alpha}^{\dagger }$ ($c_{k\alpha}$) is the creation (annihilation) operator for electrons with momentum-$k$ and spin-$\alpha $, $\xi_k = -2t(\cos k_x + \cos k_y) - 2t_z\cos k_z -\mu$ with $\mu$ the chemical potential, $v_{kk'} = -J(\eta^s_k\eta^s_{k'}+\eta^d_k\eta^d_{k'})$ with $\eta^{s,d}_k = \cos k_x \pm \cos k_y$, $p(k,q) = c_{-k+q/2\downarrow}c_{k+q/2\uparrow}$ is the pairing operator, and $N$ the total number of lattice sites. Here, we mainly consider the pairing in $d$-wave channel. By neglecting the $s$-wave-channel coupling, the interaction $v_{kk'}$ is then simply written as $v_{kk'} = -v\eta_k\eta_{k'}$ with $\eta_k = \eta^d_k$. For the $s$-wave pairing, one just puts in $\eta_k = \eta^s_k$. For studying the pairing with on site attraction, one can simply set $\eta_k = 1$. In real space, the Hamiltonian given by Eq. (1) resembles the $t-J$ model. But in the $t-J$ model, double occupation on the same site is prohibited. For taking into account of this constraint, we here adopt the approximation introduced by Baskaran, Zou, and Anderson$^{29}$ assuming the hopping integrals $t$ and $t_z$ are proportional to the hole concentration $\delta$, e.g., $t = t_0\delta$ with $t_0$ a constant. For the quasi-two-dimensional system, $t_z/t \ll 1$ is supposed. Such a model has been adopted by a number investigators for studying the $d$-wave superconductivity as well as the pseudogap phenomena in cuprates.$^{13-15,30}$ Throughout this paper, we use the units in which $\hbar = k_B = 1$.

\section {GREEN'S FUNCTIONS}

To formulate our theory, we need the Green's functions for the single particles and the pairs. In Nambu's space, the Green's function of the single particles is defined as in the textbooks by:
$$G(k,\tau-\tau')= -\langle T_{\tau} c_k(\tau)c^{\dagger}_k(\tau')\rangle \eqno(2)$$
where $T_{\tau}$ is the imaginary time-$\tau$ ordering operator, $\langle\cdots\rangle $ means a statistical average, and $c^{\dagger}_k = (c_{k\uparrow}^{\dagger },c_{-k\downarrow})$. For the Green's functions of the pairs, we firstly need to define the structured pairing operators
$$P_1(q) = {\frac{1}{\sqrt N}}\sum_k\eta_kp(k,q), \eqno(3)$$
$$P_2(q) = {\frac{1}{\sqrt N}}\sum_k\eta_kp^{\dagger}(k,-q). \eqno(4)$$
The operator $P_1(q)$ annihilates a pair of particles of symmetry as described by $\eta_k$ and of total momentum $q$, while $P_2(q)$ annihilates a pair of holes of the same symmetry and momentum $q$. The Green's functions for the pairs are defined as
$$\Pi_{\mu\nu}(q,\tau-\tau')= -\langle T_{\tau}P_{\mu}(q,\tau)P^{\dagger}_{\nu}(q,\tau')\rangle \eqno(5)$$
These functions $\Pi_{\mu\nu}$ are sometimes called as pair susceptibilities. By the ladder-diagram approximation, after Fourier transformation on the imaginary time, we obtain the Dyson's equation for $\Pi_{\mu\nu}$:
$$\Pi(q,Z_m) = \chi(q,Z_m) - v\chi(q,Z_m)\Pi(q,Z_m) \eqno(6)$$
where $\Pi(q,Z_m)$ and $\chi(q,Z_m)$ are $2\times 2$ matrices, $Z_m =i2m\pi T$ with $m$ an integer is the boson imaginary frequency, and $T$ the temperature. $\chi$ is the irreducible susceptibility. A formal solution to Eq. (6) is obtained as
$$\Pi(q,Z_m) = [1 + v\chi(q,Z_m)]^{-1}\chi(q,Z_m) \eqno(7)$$
In terms of the Green's functions of single particles, the elements of $\chi$ are given by
$$\chi_{\mu\nu}(q,Z_m) = {\frac{T}{N}}\sum_{kn}\eta^2_kG_{\mu\nu}(k_1,z_n+Z_m) G_{\bar\nu\bar\mu}(k_2,z_n), \eqno(8)
$$
where $k_{1,2} = k\pm q/2$, $z_n =i(2n+1)\pi T$ with $n$ an integer is the fermion's imaginary frequency, and the subscripts $\mu$ and $\nu$ = 1, 2 with $\bar 1 = 2$ and $\bar 2 = 1$. 

If the BEC of pairs takes place in the system, the boson Green's functions $\Pi(q,Z_m)$'s diverge at $q \to 0$ and $Z_m = 0$. This is a general property of the BEC in the boson systems without long-range interactions.$^{31}$ From Eq. (7), the divergence means that
$$\det|1 + v\chi(0,0)| = 0, \eqno(9)$$
where $\chi(0,0)$ is understood as $\lim\limits_{q\to 0}\chi(q,0)$. (In case of BEC, the zero-momentum operators $P_{\mu}(0)$ are macroscopic quantities. According to Bogoliubov, they are treated as $c$-numbers.$^{31,32}$ All the boson Green's function are defined at $q \neq 0$.) Here, a number of physical meanings of Eq. (9) need to be noted. Firstly, equation (9) is exactly the requirement for the existence of Goldstone modes in the superconducting state. Since the dispersion relation for the pairs is determined by the pole of $\Pi(q,\Omega)$, the energy of the pairs $\Omega_q$ vanishes at $q \to 0$. Therefore, the Goldstone modes are the lowest excitation states for the pairs and represent the predominant fluctuations. Secondly, at $T_c$, equation (9) reduces to the Thouless criterion for the superconducting transition.$^{33}$ 

In the Matsubara-frequency space, the Green's function of the single particles is written as
$$G(k,z_n) = [z_n - \xi_k\sigma_3 - \Sigma(k,z_n)]^{-1} \eqno(10)$$
where $z_n = z_n\hat I$ with $\hat I$ the $2 \times 2$ matrix, $\sigma$ is the Pauli matrix, and $\Sigma(k,z_n)$ is the self-energy. To express the self-energy, firstly, we note that the off-diagonal part comes from averaged boson fields of momentum $q = 0$. At the superconducting state, $\langle p(k,0)\rangle$ is a macroscopic quantity as compared with any other pair fields else. Therefore, the predominant contribution is the static mean field
$$\Sigma_{12}(k,z_n) = {\frac {1}{N}}\sum_{k'}v_{kk'}\langle p(k',0)\rangle \equiv \Delta_k .\eqno(11)$$
For our uniform system, we suppose $\Delta_k$ is real. The quantity $\Delta_k \equiv \Delta\eta_k$ should be differentiated from that of the MFT since the fluctuation effect is under consideration in the present Green's function. Secondly, for the diagonal part, we take into account of the pair fluctuation terms $q \ne 0$ of the interaction. By the ladder-diagram approximation, the diagonal part of the self-energy is given by
$$\Sigma_{\mu\mu}(k,z_n) = -{\frac{T}{N}}\sum_{qm}v^2\eta^2_{k-q/2}G_{\bar\mu\bar\mu}(k-q,z_n-Z_m)\Pi_{\mu\mu}(q,Z_m) \eqno(12)$$
Because of the divergence of $\Pi_{\mu\mu}(q,Z_m)$ at $q \to 0$ and $Z_m = 0$ as noted above, our treatment for the diagonal part of the self-energy takes into account of the predominant fluctuation effect. With the special form of the self-energy, the Green's function can be written as
$$G(k,z_n)=[z_n-\Sigma_0+(\xi_k+\Sigma_3)\sigma_3+\Delta_k\sigma_1]/Z, \eqno(13)$$
$$Z=(z_n-\Sigma_0)^2-(\xi_k+\Sigma_3)^2-\Delta^2_k, \eqno(14)$$
where $\Sigma_{0,3} = (\Sigma_{11} \pm \Sigma_{22})/2$, and the arguments ($k,z_n$) have been omitted.

It should be emphasized here that Eq. (11) is exactly consistent with Eq. (9). Any improper treatment of off-diagonal self-energy leads to violation of this consistency. To see the consistency, we note that Eq. (11) is essentially equivalent to the gap equation
$$\Delta = -{\frac{T}{N}}\sum_{kn}\eta_{k}G_{12}(k,z_n) =-{\frac{vT}{N}}\sum_{kn}\eta_{k}\Delta_{k}/Z. \eqno(15)$$
On the other hand, the left-hand side of Eq. (9) reads, 
$$(1+v\chi_{11})(1+v\chi_{22})-v^2\chi^2_{12} = (1+v\chi_{+})(1+v\chi_{-}) \eqno(16)$$
where $\chi$'s take their values at $(q,Z_m)$ = (0,0), $\chi_{11} = \chi_{22}$, and $\chi_{\pm}= \chi_{11}\pm\chi_{12}$. The factor $1+v\chi_{-}$ can be written as
$$1+v\chi_{-} = 1+{\frac{vT}{N}}\sum_{kn} \eta^2_k [G_{11}(k,z_n)G_{22}(k,z_n)-G^2_{12}(k,z_n)].\eqno(17)$$
Substituting the Green's functions given by Eq. (13) into Eq. (17), we have
$$1+v\chi_{-}=1+{\frac{vT}{N}}\sum_{kn}\eta^2_k/Z = 0, 
\eqno(18)$$
which is consistent with Eq. (15) and thereby with Eq. (11).

We need one more equation to determine the chemical potential $\mu$. This equation is for the number density of electrons,
$${\frac{2T}{N}}\sum_{kn}G_{11}(k,z_n)e^{z_n\eta} = 1-\delta \eqno(19)$$
where $\eta$ is an infinitesimal positive number. The equations (7), (8), (10)-(12) and (19) form the closed system that self-consistently determines the Green's functions.

It is a tremendous task to numerically solve these equations because many multi-dimensional integrals over momentum and the summation over Matsubara's frequency need to be computed in the iterations. However, since $\Pi(q,Z_m)$ is strongly peaked with a divergence at $q \to 0$ and $Z_m = 0$, the diagonal part of the self-energy can be approximately given by$^{15,34,35}$
$$
\Sigma_{\mu\mu}(k,z_n) \approx -v^2\eta^2_kG_{\bar\mu\bar\mu}(k,z_n){\frac{T}{N}}\sum_{qm}{'}\Pi_{\mu\mu}(q,Z_m)e^{\alpha_{\mu}Z_m\eta} \eqno(20)
$$
where $\sum{'}$ means the summation over $q$ runs small $q$, and the convergent factor $e^{\alpha_{\mu}Z_m\eta}$ with $\alpha_1 = 1$ and $\alpha_2 = -1$ has been introduced. This convergent factor comes from the fact that the Green's function $G_{\bar\mu\bar\mu}(k-q,z_n-Z_m)$ in the summation in Eq. (12) is connected with the effective interaction $v^2\Pi_{\mu\mu}(q,Z_m)$. The summations over $q$ and $m$ in Eq. (20) give rise to a constant
$$\Gamma^2 = -{\frac{T}{N}}\sum_{qm}{'}v^2\Pi_{\mu\mu}(q,Z_m)e^{\alpha_{\mu}Z_m\eta}. \eqno(21)$$
The constant $\Gamma$ is named as pseudogap parameter since at $T_c$ there remains a gap in the density of states (DOS) at the Fermi energy. Note that the constant $\Gamma^2$ is independent on the subscript $\mu$ because $\Pi_{\mu\nu}(q,Z_m) = \Pi_{\bar\nu\bar\mu}(q,-Z_m)$. $\Gamma^2$ is essentially a measure of the density of the uncondensed pairs. By such an approximation, the diagonal part of the self-energy is given by
$$\Sigma_{\mu\mu}(k,z_n) \approx \Gamma^2_kG_{\bar\mu\bar\mu}(k,z_n), \eqno(22)$$
where $\Gamma_k = \Gamma\eta_k$. 

With Eq. (22), we can get an explicit expression for the Green's function of the single particles in terms of $\Gamma_k$ and $\Delta_k$. Firstly, note that two equations from the diagonal parts of Eq.(13) form the closed system for determining the diagonal parts of $G(k,z_n)$:
$$G_0=[z_n-\Gamma^2_kG_0]/Z, \eqno(23)$$
$$G_3=[\xi_k-\Gamma^2_kG_3]/Z, \eqno(24)$$
$$Z=(z_n-\Gamma^2_kG_0)^2-(\xi_k-\Gamma^2_kG_3)^2-\Delta^2_k, \eqno(25)$$
where $G_{0,3} = (G_{11} \pm G_{22})/2$, and the arguments $k$ and $z_n$ have been omitted for briefness. From Eqs. (23) and (24), one obtains $G_0/G_3 = z_n/\xi_k$. 
Substituting this result into one of the equations, e.g., Eq. (23), we can obtain a cubic equation for $G_0$ or $G_3$. Instead of writing down such an equation for $G_0$ or $G_3$, we here introduce a function $y(k,z_n)$ by $G_0 = z_n(2-y)/3\Gamma^2_k$ so that the cubic equation for $y(k,z_n)$ and the expression for the Green's function look compact. The equation for
$y(k,z_n)$ reads,
$$y^3 - 3P y - 2Q =0 \eqno(26)$$
where $P = 1+3(\Gamma^2_k-\Delta^2_k)/(\xi^2_k-z_n^2)$ and $Q = 1+{\frac{9}{2}}(\Gamma^2_k+2\Delta^2_k)/(\xi^2_k-z_n^2)$. To match the boundary condition $G(k,z_n) \to 1/z_n$ at $n \to \infty$, we should choose the real root to Eq. (26). 
The explicit form of $y(k,z_n)$ reads
$$
y = \left\{
\begin{array}{ll}
\sqrt[3]{Q+\sqrt{D}}+\sqrt[3]{Q-\sqrt{D}},&\mbox{$D>0$}\cr
2\sqrt{P}\cos (\varphi/3),&\mbox{$D<0$}
\end{array}\right.
\eqno(27)
$$
where $D = Q^2-P^3$, and 
$$\varphi = \arccos(Q/\sqrt{P^3}). \eqno(28)$$
The final expression for $G(k,z_n)$ is  
$$G(k,z_n) = [z_n + 3\Delta_k\sigma_1/(1+y) + \xi_k\sigma_3](2-y)/3\Gamma^2_k. \eqno(29)$$

For more general purposes, we need the formula for the retarded Green's function $G(k,\omega+i\eta)$ at real frequency $\omega$. It can be obtained by the analytic continuation $z_n \to \omega+i\eta$. Firstly, note that $G(k,\omega+i\eta)$ has following property:
$$\begin{array}{rl}
{\rm Re}G_0(k,-\omega+i\eta)&=-{\rm Re}G_0(k,\omega+i\eta),\\   
{\rm Re}G_{1,3}(k,-\omega+i\eta)&={\rm Re}G_{1,3}(k,\omega+i\eta),\\     
{\rm Im}G_0(k,-\omega+i\eta)&={\rm Im}G_0(k,\omega+i\eta),\\      
{\rm Im}G_{1,3}(k,-\omega+i\eta)&=-{\rm Im}G_{1,3}(k,\omega+i\eta),\\           
\end{array}$$
Therefore, we need only to perform the analytic continuation for $\omega > 0$. Secondly, $y$ can be regarded as a function of $\Gamma_k^2/(\xi_k^2-z_n^2)|_{z_n\to\omega+i\eta} \to x+i\eta{\rm sgn}(\omega) \equiv \tilde x$ with
$$x = {\Gamma_k^2 \over \xi_k^2-\omega^2}. \eqno(30)$$
It is then enough to find out $y$ at various $x$. In terms of $\tilde x$, the related quantities can be written as $P = 1+3(1-r)\tilde x$, $Q = 1+9(r+1/2)\tilde x$ with $r = \Delta^2/\Gamma^2$. Note that $y$ can be expressed by another analytic function, $y = f + P/f$, where $f$ is given by
$$f = (Q+D^{1/2})^{1/3},\eqno(31)$$
which is definitely valid at small positive $x$ as seen from Eq. (27). For $-\infty < x < \infty$, equation (31) determines the analytic continuation. To extend the definition of $f$ in the entire region ($-\infty,\infty$), we write $D = 27a\tilde x(\tilde x-x_1)(\tilde x-x_2)$ with
$$x_{1,2} = -b/a\pm\sqrt{b^2-ar}/a, \eqno(32)$$
where $a=(r-1)^3$, $b=(8r^2+20r-1)/8$. The three branch points of $D^{1/2}$ are $x_{1,2}$ and 0, with $x_1 < 0$, $x_2|_{r<1} > 0$ and $x_2|_{r>1} < x_1$. Another quantity is $x_Q = -2/9(2r+1)$ at which $Q$ changes its sign: sgn($Q$) = sgn($x-x_Q$). Since $x_1 < x_Q$, ${\rm Re}Q$ is negative at $x < x_1$. Knowing the branch points of $D^{1/2}$ and the property of $Q$, the analytic continuation of Eq. (31) is a straightforward manipulation. Using the polar coordinates, $f = \rho\exp(i\theta)$, we give the expression for $\rho$ and $\theta$ in Tables I and II. The angle $\varphi$ in Tables I and II is given by the same functional form as by Eq. (28).

It is obvious that by setting $\Gamma_k = 0$, one gets the MFT results. In contrast to the well-defined single particles by the delta-singularity in the MFT, the single particles in the present case have finite lifetimes, and the dispersion relation is not defined. The square and cubic roots characterize the singularity of the present Green's function. This leads to the broadened peaks in the density of states (DOS) other than the sharp peaks by the MFT.$^{27}$

\section {TRANSITION TEMPERATURE}

In this section, we investigate the pairing-fluctuation effect in the transition temperature $T_c$. Because the fluctuations play important role in quasi-2D systems, $T_c$ can be substantially suppressed from the values by the MFT. 

At $T_c$, $\Delta = 0$, from Eqs. (13) and (22), we can immediately obtain a more explicit expression for the Green's function
$$G(k,z_n)=(z_n+\xi_k\sigma_3)\Bigl(1-\sqrt{1+{4\Gamma_k^2\over \xi_k^2-z_n^2}}\Bigr)/2\Gamma_k^2. \eqno(33)$$

The properties of pairs are described by the susceptibility. Firstly, we analyze the behavior of the irreducible susceptibility (at $T_c$) $\chi(q,Z_m) = \chi_0(q,Z_m)+\chi_3(q,Z_m)\sigma_3$ at small $q$ and $Z_m$. The expressions for the Pauli components are given in the Appendix. Note that $\chi_0(q,Z_m)$ and $\chi_3(q,Z_m)$ are respectively even and odd functions of $Z_m$. At $T_c$, equation (9) means $1+v\chi_0(0,0) = 0$. At small $q$ and $Z_m$, we have
$$1+v\chi_0(q,Z_m)\approx c_1q^2+c_zq_z^2+O(Z_m^2) ,\eqno(34)$$
$$v\chi_3(q,Z_m)\approx - dZ_m ,\eqno(35)$$
where $c_1$, $c_z$, and $d$ are constants, $q$ in the right-hand-side of Eq. (34) represents the in-plane components of momentum. The $q_z$ term comes from the $z$-direction motion of the particles. To the leading order, the constant $c_z$ is proportional to $(t_z/t)^2$ (see the Appendix) which is much less than $c_1$. The susceptibility $\Pi_{11}(q,Z_m)$ is then approximated by
$$v\Pi_{11}(q,Z_m)\approx -1/(c_1q^2+c_zq_z^2-dZ_m).\eqno(36)$$
The pole of $\Pi_{11}(q,\Omega)$ gives rise to the low-lying excitation energy $\Omega_q = (c_1q^2+c_zq_z^2)/d$ for the pairs.  The constants $d/2c_1$ and $d/2c_z$ represent the effective masses of in-plane and out-plane motions, respectively. 

According to Eq. (21), the pseudogap parameter $\Gamma$ is calculated via
$$\begin{array}[b]{rl}
\Gamma^2 &= -{\frac{T}{N}}\sum\limits_{qm}{'}v^2\Pi_{11}(q,Z_m)e^{Z_m\eta}\cr\noalign{\vskip 3mm}
&= {\frac{Tv}{Nd}}\sum\limits_{qm}{'}{1 \over \Omega_q-Z_m}e^{Z_m\eta}\cr\noalign{\vskip 3mm}
&= {\frac{v}{Nd}}\sum\limits_{q}{'}B(\Omega_q),
\end{array}  \eqno (37)$$
where $B(\Omega_q)= 1/[\exp(\Omega_q/T)-1]$ is the Bose function. Equation (37) clearly reflects the fact that the pairs are independent on each other. So the total number of the pairs is the summation of the boson-occupation numbers on each state. A character of the independent pairs (single pairs) is $\Omega_q \propto q^2$ at $q_z = 0$. The $q$-integral in Eq. (37) is over a region of small momentum. Since $\Omega_q$ depends weakly on the out-plane wave number $q_z$, the integral over $q_z$ can be taken in the range $(-\pi,\pi)$. Though the $z$-direction dispersion is weak, it prevents the summation over $q$ from a logarithm divergence at the $q = 0$ limit and ensures a finite transition temperature $T_c$. This is in agreement with the MWH theorem. The cutoff $q_c$ for the in-plane wave number is determined such that the largest in-plane energy $\Omega_{q_{c}} = 2\Gamma$ since a pair that is a bound state of two particles is meaningful only within the gap. Note that $\Omega_q$ is a two-particle excitation energy. The largest gap for a pair of particles should be 4$\Gamma$. But regarding the $d$-wave pairing, the average gap is less than 4$\Gamma$. The cutoff $2\Gamma$ is a reasonable choice. We have examined the insensitive dependence of $T_c$ and $\Gamma$ on this cutoff. Even by putting in $q_c = \pi$, the results for $T_c$ and $\Gamma$ change very little.

Equations (9), (19) and (37) self-consistently determine the quantities $T_c$, $\mu$, and $\Gamma$. The result for $T_c$ as a function of hole concentration $\delta$ (solid line) is plotted in Fig. 1. In the numerical calculation, we take $v/2t_0 \simeq 0.1$ and $t_z/t \simeq 0.01$ for describing the cuprates.$^{14}$ For $\rm La_2CuO$, $v \simeq 0.13$ eV has been determined by experiments.$^{36}$ Therefore, the choice of $v/2t_0$ corresponds to $t_0 \simeq 0.65$ eV, which is consistent with estimates from experiment data.$^{37}$ The maximum transition temperature $T_{c,Max} \approx 0.01575 t_0$ ($\approx$ 118 K) obtained by the present theory appears at a certain $\delta$ between 0.125 and 0.15. The MFT result and the experiment data$^{38}$ are also shown in Fig. 1 for comparison. The phase boundary of superconductivity given by the present theory is close to the experiment results. In the optimally to over doped region, the present theory fits the experiment data very well. The improvement of the present theory to the MFT is significant. Because of the fluctuations, $T_c$ is substantially suppressed from that of the MFT. Especially in the under doped region, contrary to the MFT, $T_c$ of the present calculation decreases with decreasing $\delta$ and vanishes at $\delta$ = 0. $T_c$ as a function of $\delta$ by the present calculation shows a parabolic behavior the similar as the experiment data, while the MFT result is a monotonically decreasing function of $\delta$. 

By the MFT, at $T = T_c < T_{c, {\rm MFT}}$, there are pairs in the condensate. When pairing fluctuations set in the system, all these pairs are changed to be incoherent pairs. This results in the reduction of $T_c$. This fluctuation effect is more notable at the strong coupling regime where the strength $v/t_0\delta$ is large; the particles intend to pair independently. This is clearly reflected by the pseudogap parameter $\Gamma$ shown in Fig. 1 as the dashed line.

One may ask why $T_{c, MFT} \ne 0$ at $\delta$ = 0. This is because there exists local pairing at $\delta$ = 0 by the MFT. We can analytically show that $T_{c, MFT}$ is finite at $\delta$ = 0. Consider the gap equation at $T_c$
$${v\over 2N}\sum_k\eta_k^2\tanh(\xi_k/2T_c)/\xi_k = 1.$$
At $\delta$ = 0, we have $\xi_k$ = 0, and thereby $\tanh(\xi_k/2T_c)/\xi_k \to 1/2T_c$. One then obtain
$$T_c = {v\over 4N}\sum_k\eta_k^2 = {v\over 4}.$$
For our system, $v$ = 130 meV, $T_{c,Max} \approx 0.01575 t_0$ = 10.2 meV, therefore $v/4T_{c,Max} \approx 3.2$. At small $\delta$, $\xi_k$ is a small quantity, and $\tanh(\xi_k/2T_c)/\xi_k$ can be expanded as $\tanh(\xi_k/2T_c)/\xi_k \approx 1/2T_c - \xi_k^2/12T_c^3$. The equation for $T_c$ reads
$$T_c \approx {v\over 4}-{v\over 24T_c^2N}\sum_k\eta_k^2\xi_k^2.$$
Apparently, $T_c < v/4$ at small $\delta$.

In passing, we here explain the meaning of pseudogap parameter using the Green's function given by Eq. (33). We consider the spectral function at $T_c$, $A(k,E) = -{\rm Im} G_{11}(k,E+i\eta)/\pi = \sqrt{(\xi_k^2+4\Gamma_k^2-E^2)/(E^2-\xi_k^2)}(E+\xi_k)/2\pi\Gamma^2_k$, which is nonzero only for $E^2-4\Gamma^2_k < \xi_k^2 < E^2$. The noninteracting delta-function peak becomes a square root singularity. Near the Fermi energy, the $k$-space is constrained so that the volume decrease at $E \to 0$, resulting in a suppression of DOS at the Fermi energy. This leads to the formation of a pseudogap in the DOS.$^{27}$ 

\section {COLLECTIVE MODES AND PSEUDOGAP AT THE SUPERCONDUCTING STATE}

At the superconducting state, the properties of the pairs are qualitatively different from that we have discussed in the previous section. The expression for the pseudogap parameter will be also different from Eq. (37). Below the transition temperature, with the BEC taking place, the uncondensed pairs become moving collectively.

To investigate the properties of the pairs, we start from analyzing the susceptibilities. The low-energy pairs of particles are described by $\Pi_{11}(q,Z_m)$ at small $q$ and $Z_m$. This function is given via
$$v\Pi_{11} = 1- [1+v(\chi_0-\chi_3)]/D, \eqno(38)$$
$$D = (1+v\chi_{-})(1+v\chi_{+})-v^2\chi_3^2, \eqno(39)$$
where $\chi_{\pm} = \chi_0\pm \chi_1$, the arguments $(q,Z_m)$ have been omitted for briefness. (The denominator $D$ should not be confused with the quantity $D$ in Eq. (27).) Since $1+v\chi_{-}(0,0) = 0$ as shown by Eq. (18), at $q \to 0$ and $Z_m \to 0$, we have  
$$1+v\chi_{-}(q,Z_m) \approx c_1q^2 +c_zq_z^2-c_0Z_m^2, \eqno(40)$$
$$v\chi_{3}(q,Z_m) \approx -dZ_m. \eqno(41)$$
The constants $c_1$ and $c_z$ are different from that we obtained in Eq. (34) because of a contribution from $\chi_1$ in the present case. Note that $1 + v\chi_{+} = 1 + v\chi_{-}+ 2v\chi_{1}$, we get
$$1+v\chi_{+}(q,Z_m) \approx 2p+c'_1q^2 +c'_zq_z^2, \eqno(42)$$
where $p \equiv v\chi_{1}(0,0)$. For the denominator $D$, we obtain
$$\begin{array}[b]{rl}
D&=(c_1q^2 +c_zq_z^2)(2p+c'_1q^2 +c'_zq_z^2)-(2pc_0+d^2)Z_m^2\cr\noalign{\vskip 3mm}
& \equiv (2pc_0+d^2)(\Omega_q^2-Z_m^2),
\end{array} \eqno(43)$$
which defines the excitation energy $\Omega_q$ of the pairs. At very small $q$, $\Omega_q \approx \sqrt {2p(c_1q^2 +c_zq_z^2)/u}$ with $u = 2pc_0+d^2$. At $q_z$ =0, $\Omega_q\propto q$ at $q \to 0$. This is one of the features of the collective modes, in contrast to that of the single pairs $\Omega_q\propto q^2$. 
By the same consideration, one can approximate $1+v\chi_{0}$ as
$$1+v\chi_{0}(q,Z_m) \approx p+c^{0}_1q^2 +c^{0}_zq_z^2. \eqno(44)$$
The function $v\Pi_{11}(q,Z_m)$ is then approximated by
$$v\Pi_{11}(q,Z_m) \approx {1\over u}{p+c^{0}_q+dZ_m \over Z_m^2-\Omega_q^2}, \eqno(45)$$
where $c^{0}_q = c^{0}_1q^2 +c^{0}_zq_z^2$. Note that at $T_c$, $p$ = 0, $c^{0}_1 = c'_1 = c_1$, and $c^{0}_z = c'_z = c_z$. We keep the small terms besides the constant terms in Eqs. (42) and (44) in order to formally reproduce the same formula for $\Pi_{11}$ as at $T_c$.

Similarly, we can obtain the formulae for $\Pi_{12}$, $\Pi_{21}$, and $\Pi_{22}$. The form for the matrix $\Pi$ then is
$$\Pi (q,Z_m) = {M_{+}(q)\over Z_m-\Omega_q} - {M_{-}(q)\over Z_m+\Omega_q}, \eqno(46)$$
where $M_{\pm}(q)$ are two matrices. To the leading order $O(1/\Omega_q)$, we have 
$$M_{+}(q) = M_{-}(q) = {p\over 2uv\Omega_q}(1-\sigma_1). \eqno(47)$$
Equation (46) reminds us that the pairing operators $P^{\dagger}(q) \equiv [P^{\dagger}_1(q),P^{\dagger}_2(q)]$ can be represented by two eigenmodes $C_1(q)$, and $C_2(q)$:
$$P(q) = A_qC_1(q)+B_qC^{\dagger}_2(q), \eqno(48)$$
where $A_q$ and $B_q$ are two spinors, determined by
$A_qA^{\dagger}_q = M_{+}(q)$, and $B_qB^{\dagger}_q = M_{-}(q)$. To the leading order $O(1/\sqrt{\Omega_q)}$, we get
$$A^{\dagger}_q = B^{\dagger}_q = \sqrt{p\over 2uv\Omega_q}(1,-1). \eqno(49)$$ 
These two eigenmodes are the collective modes, with the same energy $\Omega_q$.

By applying Eq. (45) to Eq. (21), the pseudogap parameter $\Gamma$ is now calculated via
$$\Gamma^2 = -{\frac{Tv}{Nu}}\sum_{qm}{'}{p+c^{0}_q+dZ_m \over Z_m^2-\Omega_q^2}e^{Z_m\eta}.\eqno(50)$$
Carrying out the Matsubara sum, we have
$$\Gamma^2 = {\frac{Tv}{Nu}}\sum_{q}{'}\{[B(\Omega_q)+{1\over 2}]{p+c^{0}_q \over \Omega_q}-d/2\}, \eqno(51)$$
where the cutoff $q_c$ for the in-plane wave number is now determined by $\Omega_{q_{c}} = 2\sqrt{\Delta^2+\Gamma^2}$ by the similar reason as for Eq. (37). The summation over $q$ mainly comes from the first term $[B(\Omega_q)+{1\over 2}]p/\Omega_q$ because which diverges like $q^{-2}$ at $q \to 0$. Clearly, equation (51) is different from Eq. (37). The total number of pairs cannot be written as the summation over the Bose distributions because the pairs move collectively. We note that besides the boson occupation number $B(\Omega_q)$, the term $1/2$ represents the contribution from the zero-point motion. This is another feature of the collective modes. At the ground state, $B(\Omega_q)=0$, the integrand in Eq. (51) behaves like $1/q$ at $q \to 0$. Two conclusions are deduced from this fact: (1) $\Gamma^2$ is finite at the ground state. (2) Broken-symmetry state may survive in a pure 2D system at $T = 0$.  This is in agreement with the perturbation theories.$^{20,23}$ The numerical results for $\Gamma$ have been shown in Ref. 27

Why can the collective modes appear below $T_c$? Consider the case of generating a pair of particles of total momentum $q$ in the system. The generation process influences the condensate. This pair comes from a superposition of the fluctuation of all the pairs in the condensate. This is not the case at $T \geq T_c$. There is no condensation above $T_c$. Each pair can be constructed only from two single particles. So the pairs with different momentum are independent with each other.

\section {CONCLUSION AND REMARKS}

In summary, we have studied the pairing fluctuation effects in the quasi-2D superconducting system. The fluctuations over the mean-field pairing come predominately from the Goldstone modes. We treat these collective modes and the single particles on the equal-footing manner. The respective fermion and boson Green's functions for the single particles and the pairs are self-consistently determined by a number of integral equations. The single-particle Green's function is given explicitly by Eqs. (27) and (29). The pairing fluctuations result in lifetimes for the single particles. 

We have investigated the superconductivity in cuprates using the tight-binding model with $d$-wave attraction. $T_c$ is substantially suppressed from its MFT value. The phase boundary of superconducting state given by the present theory can reasonably describe the experiment results. Also, the pairing fluctuations are relevant with the pseudogap physics in cuprates. 

The form of the Green's function given by Eq. (33) may be still meaningful at $T$ slightly above $T_c$. It is obtained by observing that the long-range fluctuations at $Z_m = 0$ have divergently contribution to the self-energy. At $T \geq T_c$, though there is no such a divergence, the contribution comes from the long-range fluctuations at small $Z_m$ is still large. The experimental observations indicate that the pseudogap parameter weakly depends on $T$ slightly above $T_c$.$^{26}$ Therefore, the Green's function obtained at $T_c$ may be useful for studying the properties of the pseudogap state near $T_c$.

For the phase boundary, there is still obvious discrepancy between the present theory and the experiment at very under doped region. This may date from the crude treatment of the short-range pair correlations. Local pairing without long-range phase coherence is not fully taken into account in the present model. Besides this, the short-range antiferromagnetic coupling is not correctly counted in. To describe the antiferromegnetism in cuprates at very small $\delta$, one needs to restart with the $t-J$ model.

\vskip 4mm
\centerline {\bf ACKNOWLEDGMENTS}
\vskip 2mm

This work was supported by Natural Science Foundation of China (GN. 10174092) and by Department of Science and Technology of China (G1999064509).

\vskip 4mm
\centerline {\bf APPENDIX}
\vskip 2mm

In this Appendix, we give the expressions for the Pauli components of the irreducible susceptibility $\chi=\chi_0+\chi_1\sigma_1+\chi_3\sigma_3$. Then, we derive the equation for the coefficient $c_z$ describing the $z$-direction dispersion relation of the pairs. 

The Pauli components of $\chi$ are given by:
$$\chi_0(q,Z_m) = {\frac{T}{N}}\sum_{kn}\eta^2_k[G_0(k_1,z_n^{+})G_0(k_2,z_n)
-G_3(k_1,z_n^{+})G_3(k_2,z_n)], \eqno(A1)$$
$$\chi_1(q,Z_m) = {\frac{T}{N}}\sum_{kn}\eta^2_kG_1(k_2,z_n^{+})G_1(k_1,z_n), \eqno(A2)$$
$$\chi_3(q,Z_m) = {\frac{T}{N}}\sum_{kn}\eta^2_kG_3(k_1,z_n)[G_0(k_2,z_n^{-})-G_0(k_2,z_n^{+})], \eqno(A3)$$
where $k_{1,2} = k\pm q/2$, and $z_n^{\pm} = z_n\pm Z_m$. Apparently, $\chi_0(q,Z_m)$ and $\chi_1(q,Z_m)$ are even about $Z_m \to -Z_m$, while $\chi_3(q,Z_m)$ is odd. 

The quantity $c_z$ in Eq. (40) is defined as
$$\begin{array}[b]{rl}
c_z&={v\over 2}{\partial^2\over \partial q^2_z}\chi_{-}(q_z,0)|_{q_z=0}\cr\noalign{\vskip 3mm}
&=-{\frac{Tv}{2N}}\sum\limits_{kn\mu}\eta^2_kG_{\mu}(k,z_n){\partial^2\over \partial k^2_z}G_{\mu}(k,z_n)s_{\mu},
\end{array} \eqno(A4)$$
where $\mu$ = 0, 1, and 3, $s_0 = -1$, $s_1 = s_3 = 1$. For our quasi-2D system, the dependence of the Green's functions on $k_z$ is only through $\xi_k = -2t(\cos k_x + \cos k_y) - 2t_z\cos k_z -\mu$, the derivative of the Green's functions in Eq. (A4) can be calculated by
$${\partial^2\over \partial k^2_z }G_{\mu}(k,z_n) =
4t^2_z\sin^2k_z{\partial^2\over \partial \xi^2_k} G_{\mu}(k,z_n)
+2t_z\cos k_z{\partial\over \partial\xi_k }G_{\mu}(k,z_n). 
\eqno(A5)$$
Moreover, $t_z/t \ll 1$, to the first order of $t_z/t$, we have
$$G_{\mu}(k,z_n)=\bar G_{\mu}(k,z_n)- 2t_z\cos k_z{\partial\over \partial \xi_k} \bar G_{\mu}(k,z_n), \eqno(A6)$$
where the bar functions mean that in which $t_z$ is set to 0. Substituting (A5) and (A6) into (A4) and carring out the $k_z$-integral, we finally get
$$c_z ={\frac{Tvt^2_z}{N}}\sum_{kn\mu}\eta^2_k[{\partial\over \partial\xi_k}\bar G_{\mu}(k,z_n)]^2s_{\mu}. \eqno(A7)$$
The $k$-summation in Eq. (A7) is essential a two-dimensional in-plane $k$-integral. 

By the similar procedures, we can get the equations for $c'_z$ appeared in Eq. (42) and $c^0_z$ in Eq. (44). The equation for 
$c'_z$ is obtained from Eq. (A7) by setting $s_1 = -1$. For 
$c^0_z$, we set $s_1 = 0$.

\begin{table}
\caption{Function $f(x+i\eta)=\rho\exp(i\theta)$ at $r < 1$. }
\begin{tabular}{cccc}
$\rho$               & $\theta$    &    $x$               & Arg($D$)\\ \hline
$|Q-\sqrt{D}|^{1/3}$ & $\pi/3$  &  $(-\infty,x_1)$        & $2\pi$  \\
$\sqrt{P}$           & $\varphi/3$ & $(x_1,0)$            & $\pi$   \\
$|Q+\sqrt{D}|^{1/3}$ & 0           & $(0,x_2)$            & 0       \\
$\sqrt{P}$           & $-\varphi/3$& $(x_2,\infty)$       & $-\pi$  \\
\end{tabular}
\end{table}

\begin{table}
\caption{Function $f(x+i\eta)=\rho\exp(i\theta)$ at $r > 1$. }
\begin{tabular}{cccc}
$\rho$               & $\theta$          &    $x$               & Arg($D$)\\ \hline
$\sqrt{P}$           & $(2\pi-\varphi)/3$& $(-\infty,x_2)$      & $3\pi$  \\
$|Q-\sqrt{D}|^{1/3}$ & $\pi/3$           & $(x_2,x_1)$          & $2\pi$  \\
$\sqrt{P}$           & $\varphi/3$       & $(x_1,0)$            & $\pi$   \\
$|Q+\sqrt{D}|^{1/3}$ & 0                 & $(0,\infty)$         & 0       \\
\end{tabular}
\end{table}

\vskip 3mm
\begin{figure}[tbp]
\centerline{\epsfig{file=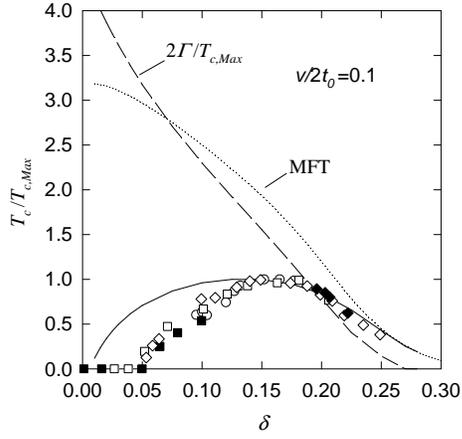,width=6 cm}}
\vskip 2mm
\caption{Transition temperature $T_c$ as a function of hole concentration $\delta$. The solid and dotted lines represent the results of present approach and the MFT, respectively. Both theoretical calculations use the same normalization constant $T_{c,Max}$ = 118 K. The symbols indicate the experiment data for cuprates:$^{38}$ $\rm Y_{1-x}Ca_xBa_2Cu_3O_6$ (solid squares), $\rm Y_{0.9}Ca_{0.1}Ba_2Cu_3O_{7-y}$ (open squares), $\rm La_{2-x}Sr_xCuO_4$ (open diamonds), $\rm Y_{1-x}Ca_xBa_2Cu_3O_{6.96}$ (solid diamonds), and $\rm YBa_2Cu_3O_{7-y}$ (open circles). The result for the pseudogap parameter $\Gamma$ is shown as the dashed line.}
\end{figure}


\begin{references}
\bibitem{[28]} In real space, the Hamiltonian reads
$$H = \sum_{<ij>}t_{ij}c_{i\alpha }^{\dagger }c_{j\alpha }+{v\over 2}\sum_{<ij>}({\vec S_i}\cdot{\vec S_j} -n_in_j/4)$$
where the $<ij>$-summations run over the nearest-neighbor (NN) sites, but the second one runs over the NN sites in the same plane. We neglect the small coupling between electrons in different planes.
\end{references}
\end{document}